\documentclass[11pt,a4paper]{article}
\pdfoutput=1
\usepackage{jheppub}	
\usepackage{graphicx}				
\usepackage{amsmath}
\usepackage{color}
\usepackage{mathtools}
\usepackage{amssymb}
\usepackage{booktabs}
\usepackage{comment}
\usepackage{mathrsfs}
\usepackage{enumerate}

\usepackage{hyperref}

\hypersetup{
    bookmarks=true,
    colorlinks,
    citecolor=blue,
    filecolor=black,
    linkcolor=blue,
    urlcolor=blue
}

\usepackage{overpic}
\usepackage[caption=false]{subfig}
\usepackage{array}
\usepackage{rotating}

\title{Holographic Complexity: Stretching the Horizon of an Evaporating Black Hole}

                                           \author{Lukas Schneiderbauer,}
                                           \author{Watse Sybesma}
                                           \author{and L\'{a}rus Thorlacius}
                                           \affiliation{Science Institute,\\
                                           University of Iceland, \\Dunhaga 3, 107 Reykjav\'{i}k, Iceland.}
                                           \emailAdd{lukas.schneiderbauer@gmail.com}
                                           \emailAdd{watse@hi.is}
                                           \emailAdd{lth@hi.is}
\abstract{
We obtain the holographic complexity of an evaporating black hole in the semi-classical RST model of
two-dimensional dilaton gravity, using a volume prescription that takes into account the higher-dimensional
origin of the model. For classical black holes, we recover the expected late time behaviour of the complexity,
but new features arise at the semi-classical level. By considering the volume inside the stretched 
horizon of the evolving black hole, we obtain sensible results for the rate of growth of the complexity, 
with an early onset of order the black hole scrambling time followed by an extended period where the
rate of growth tracks the shrinking area of the stretched horizon as the black hole evaporates.
  }
                                           
\begin{document}
                                           \maketitle

\section{Introduction}
\label{sec:intro}

The interior of a black hole is the archetype of an emergent spacetime in the holographic approach
to quantum gravity. The principle of black hole complementarity posits that the interior geometry and
any matter that enters a black hole can be described in terms of a finite number of quantum mechanical
degrees of freedom associated with a stretched horizon located outside the event horizon \cite{Susskind:1993if}.
In order to reproduce black hole thermodynamics, the number of stretched horizon degrees of freedom
should match the Bekenstein-Hawking entropy and the dynamics must be sufficiently chaotic to scramble
quantum information on a relatively short timescale, but, beyond that, the precise nature of the stretched
horizon dynamics and the holographic encoding of the black hole interior remain elusive. In what follows,
we will not make any specific assumptions about the scrambling dynamics but it can be useful to keep in
mind a collection of qubits undergoing k-local interactions as a simple model \cite{Susskind:2018pmk}.

Quantum complexity has in recent years emerged as an important entry in the holographic dictionary
for black holes following Susskind's conjecture that the expanding spatial volume of the Einstein-Rosen
bridge of a two-sided eternal black hole reflects the growing complexity of a corresponding quantum
state \cite{Susskind:2014rva}.
In the present paper, we explore the relation between quantum complexity and interior
black hole geometry in the context of semi-classical black holes that are formed by gravitational
collapse and subsequently evaporate due to the emission of Hawking radiation. Our main result,
based on explicit calculations in a two-dimensional dilaton gravity model that allows analytic study of
semi-classical effects, is that the rate of growth of holographic complexity precisely tracks the
shrinking area of the stretched horizon as the black hole evaporates, where the stretched horizon
is taken to be a membrane with an area larger than that of the event horizon by
order one in the appropriate units of the model.

The dilaton gravity model has explicit classical solutions describing black hole formation
from arbitrary incoming matter energy flux. We focus, for simplicity, on black holes formed by an infalling
thin shell and start off by adapting the complexity as volume conjecture to this context. A suitably defined
volume functional exhibits precisely the expected linear growth with time at late times and by restricting to
the volume inside the stretched horizon of the dynamically formed black hole one finds reasonable
early-time behaviour as well. We then consider a semi-classical extension of the model where the field
equations remain analytically soluble and numerically evaluate the volume functional in an evaporating
black hole background.

The transitory nature of semi-classical black holes highlights certain technical aspects of 
the identification between complexity and volume, that can often be ignored when considering
classical black holes. In the present paper, we only consider the volume representation of the semi-classical 
black hole complexity, where these issues are relatively easy to address. The alternative formulation of holographic 
complexity in terms of the action on a Wheeler-DeWitt patch \cite{Brown:2015bva,Brown:2015lvg} is also of interest 
for these dilaton gravity models, but it is more subtle to implement at the semi-classical level, and we postpone 
this to a forthcoming paper \cite{sst_inprep}.  

%

\section{Complexity of classical CGHS black holes}
\label{sec:classical}
We work within a class of two-dimensional dilaton gravity theories first introduced by Callan, Giddings,
Harvey, and Strominger (CGHS) \cite{Callan:1992rs}. These are simple toy theories for black hole physics
that can be systematically studied at the semi-classical level. They have classical solutions that describe
black hole geometries with a spacelike singularity inside an event horizon. The black holes include static
two-sided black holes and also dynamical black holes formed by the gravitational collapse of matter fields.
The quantization of matter fields in a black hole background leads to Hawking radiation and its 
back-reaction on the geometry causes the black hole to evaporate. The subsequent evolution is 
particularly simple to track in a variant of the semi-classical model that was introduced by Russo, Susskind, 
and Thorlacius (RST), where the semi-classical field equations can be solved analytically~\cite{Russo:1992ht}. 

The original CGHS model can be viewed as a
spherical reduction of a four-dimensional dilaton gravity theory in a near-extremal magnetically charged
black hole background \cite{Giddings:1992kn,Banks:1992ba}. The two-dimensional theory captures the
low-energy dynamics of radial modes in the near-horizon region of higher-dimensional geometry. The
volume that is to be identified as the quantum complexity is that of a spacelike three-dimensional surface
in the original theory, rather than the length of a spacelike curve in two-dimensions, and this will be
reflected in our calculations below.

The CGHS model and related semi-classical models were studied
extensively in the early 1990's and several reviews were written at that time,
including \cite{Harvey:1992xk,Thorlacius:1994ip,Strominger:1994tn,Giddings:1994pj}.
We will be brief and only introduce the minimal ingredients needed for the purposes of this paper.

The classical CGHS action,
\begin{equation}\label{eq:CGHS}
	S_{\text{CGHS}}
	=
	\int_{\mathcal{M}} d^{2}x\sqrt{-g}
	\left[
		e^{-2\phi}
		\left(
			R
			+
			4(\nabla\phi)^{2}
			+
			4\lambda^{2}
		\right)
		-
		\frac{1}{2}
		\sum^{N}_{i=1}(\nabla f_{i})^{2}
	\right]
	\,,
\end{equation}
involves the two-dimensional metric, a scalar dilaton field, and matter in the form of $N$ minimally coupled scalar 
fields $f_i$. The two-dimensional theory inherits a scale $\lambda$ from the parent theory set by the magnetic
charge of the near-extremal black hole.  In the following, we take length to be measured in units of $\lambda^{-1}$
and thus set $\lambda=1$. 

We find it convenient to work in a conformal gauge,
\begin{equation}
{ds^{2}=-e^{2\rho}dx^{+}dx^{-}},
\end{equation}
and use a residual conformal reparametrisation to choose coordinates where $\rho=\phi$. These are referred to 
as Kruskal coordinates for reasons that will become apparent below. In this coordinate system, the classical 
equations of motion and constraints reduce to
\begin{equation}\label{eq:CGHSeom1}
	\partial_{+}\partial_{-}f_{i}=0
	\,,
	\qquad
	\partial_{+}\partial_{-}e^{-2\phi}
	=
	-1
	\,,
	\qquad
	\partial_{\pm}^{2}e^{-2\phi}
	=
	-T^{f}_{\pm\pm}
	\,,
\end{equation}
where $T_{\mu\nu}^{f}$ is the energy-momentum tensor of the $f_{i}$ matter fields.

In order to obtain the holographic complexity of a classical CGHS black hole, we introduce a volume functional 
\begin{equation}\label{eq:volume_prop_mod}
	V = \int ds \, e^{-2\phi} \sqrt{ g_{\mu \nu} \dot{y}^{\mu} \dot{y}^\nu }
	\,,
\end{equation}
where $y^\mu(s)$ is a spacelike curve in the two-dimensional spacetime and the 
integrand includes a factor of $e^{-2\phi}$, which is proportional to the area of the local transverse 
two-sphere $S^2$ of the near-extremal dilaton black hole (in Einstein frame) in the higher-dimensional parent theory.
A corresponding factor was included when defining $\mathcal{C}_V$ for black holes in two-dimensional 
Jackiw-Teitelboim gravity in~\cite{Brown:2018bms}. In the case of the Jackiw-Teitelboim black hole, the transverse 
$S^2$ has constant area and the calculation of the complexity reduces to calculating a two-dimensional geodesic length. 
For a CGHS black hole, on the other hand, the transverse area depends on spatial location and curves that 
maximize \eqref{eq:volume_prop_mod} are {\it not\/} geodesics. 

\subsection{Complexity of a two-sided black hole}
\label{subsec:eternal}
\begin{figure}[h]
\vspace{0.5cm}
	\begin{center}
		\begin{overpic}[width=.7\textwidth]{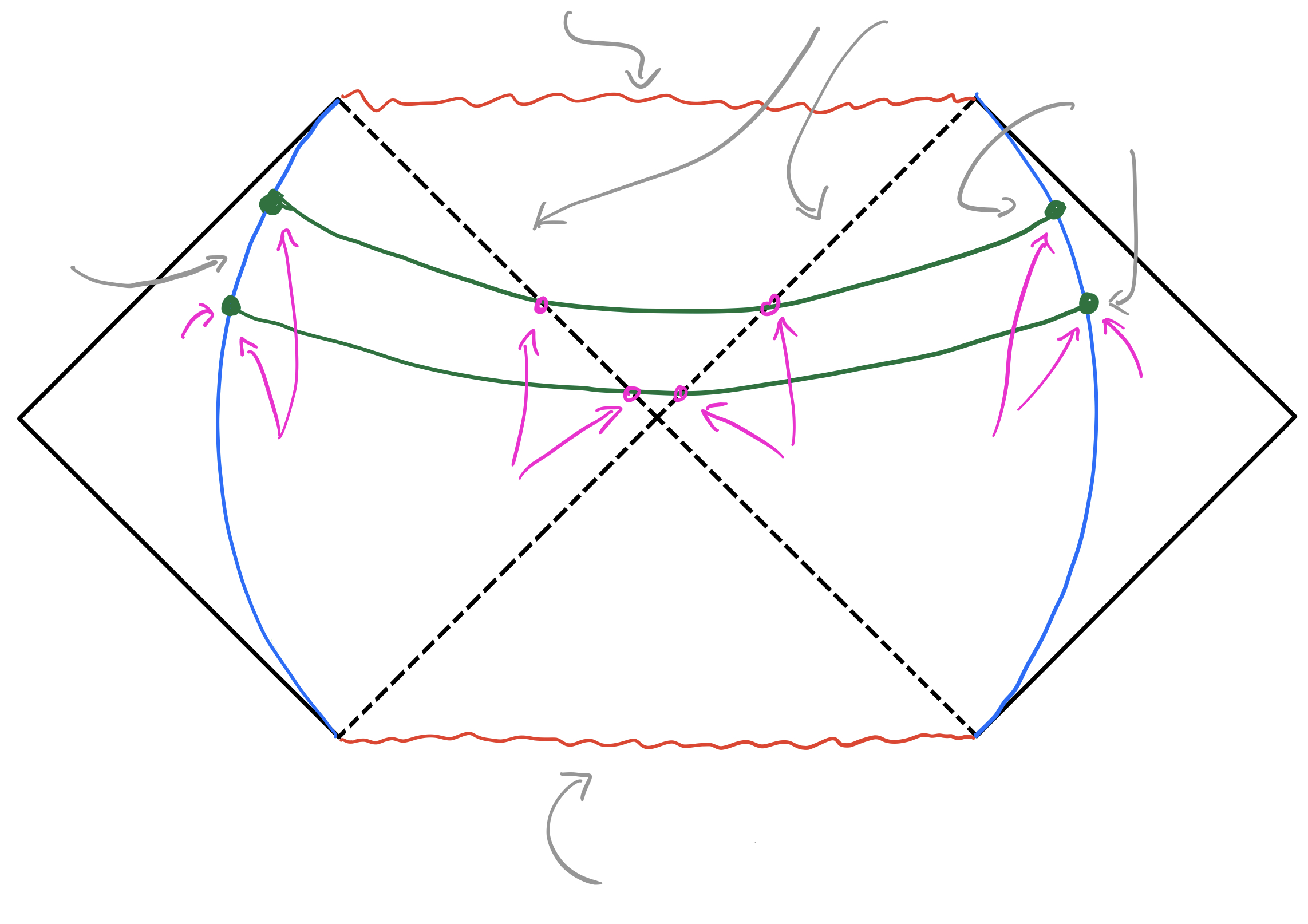}
			\put (-8.5,52.5) {\footnotesize{or Anchor Curve}}
			\put (-9,56) {\footnotesize{Stretched Horizon}}	
			\put (54,70) {\footnotesize{Event Horizons}}		
			\put (12,68) {\footnotesize{Black Hole Singularity}}
			\put (82,63) {\footnotesize{Anchor}}
			\put (83,60) {\footnotesize{Points}}
			\put (48,7) {\footnotesize{White Hole}}
			\put (48.5,4) {\footnotesize{Singularity}}
			\put (18,35) {\footnotesize{$\tau=\tau_{L}$}}	
			\put (35,32) {\footnotesize{$\tau=0$}}		
			\put (58,33) {\footnotesize{$\tau=\mu$}}		
			\put (73,35) {\footnotesize{$\tau=\tau_{R}$}}		
			\put (85,39) {\footnotesize{$(\sigma_{a},t_{R})$}}
			\put (6,41) {\footnotesize{$(\sigma_{a},t_{L})$}}	
			\put (75,11.5) {\footnotesize{$i^{-}$}}	
			\put (24,11.5) {\footnotesize{$i^{-}$}}	
			\put (24,64) {\footnotesize{$i^{+}$}}	
			\put (75,64) {\footnotesize{$i^{+}$}}	
			\put (100,38) {\footnotesize{$i^{0}$}}	
			\put (-1.5,37.5) {\footnotesize{$i^{0}$}}	
			\put (9.5,24.5) {\footnotesize{$\mathcal{I}^{-}$}}	
			\put (87.5,24.5) {\footnotesize{$\mathcal{I}^{-}$}}
			\put (87.5,51.5) {\footnotesize{$\mathcal{I}^{+}$}}	
		\end{overpic}
	\end{center}
\vspace{-0.7cm}
\caption{
This cartoon depicts the Penrose diagram of a two-sided eternal CGHS black hole. 
}\label{fig:eternal}
\end{figure}
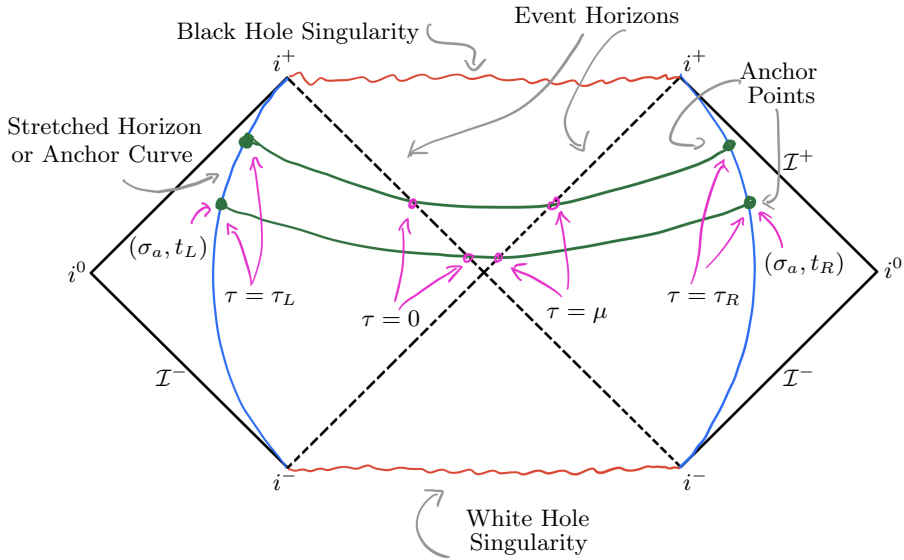

The first solution we consider describes a two-sided eternal black hole, 
\begin{equation}\label{eq:CGHSeternal}
	e^{-2\phi}=e^{-2\rho}=M-x^{+}x^{-}
	\,,
	\qquad
	f_i=0 \,.
\end{equation}
From the Ricci scalar, 
\begin{equation}\label{eq:CGHSricci}
	R
	=
	-2\nabla^{2}\rho
	=
	\frac{4M}{M-x^{+}x^{-}}
	\,,
\end{equation}
it is apparent that, for $M>0$, the curvature is singular on the spacelike curves $x^{+}x^{-}=M$,
corresponding to the white hole and black hole singularity. 
The Penrose diagram, shown in Figure~\ref{fig:eternal}, is identical to the one obtained
for a Schwarzschild black hole in 3+1 dimensional Einstein gravity.  
The event horizon is at $x^{+}x^{-}=0$ and there are two asymptotic regions, $-x^{+}x^{-}\to\infty$, 
where the curvature goes to zero. 
One can introduce Schwarzschild-like coordinates $(t,\sigma)$ in the outside region on the right
(where $x^+>0$ and $x^-<0$) via the transformation $x^\pm = \pm e^{\pm t+\sigma}$. 
The corresponding coordinate transformation on the left (where $x^+<0$ and $x^->0$) is 
given by $x^\pm = \mp e^{\mp t+\sigma}$. 
With these conventions, the metric in $(t,\sigma)$ coordinates approaches the two-dimensional 
Minkowski metric as $\sigma\to +\infty$ on both sides of the black hole and $t$ propagates to
the future in the `upward' direction on both sides.

The volume in \eqref{eq:volume_prop_mod} is divergent for spacelike curves that extend all the 
way to spatial infinity. In order to obtain a finite expression for the complexity, we introduce timelike
anchor curves outside the black hole where the volume integral is cut off (see Figure~\ref{fig:eternal}). 
We find it convenient to use anchor curves on which the dilaton field is constant, $\phi(x_a^+,x_a^-)=\phi_a$, 
providing a coordinate invariant notion of spatial position outside the black hole, and place the curves
symmetrically on the left- and right-hand side of the black hole. The anchor curves take a particularly
simple form in the Schwarzschild-like coordinates, where they are curves of constant $\sigma=\sigma_a$, with
\begin{equation}\label{eq:sigmaa}
\sigma_a=\frac12\log\left(e^{-2\phi_a}-M\right)\,,
\end{equation}
where the anchor curve on the left (right) is parametrised by $t_L$ ($t_{R}$).

Our prescription for the volume complexity $\mathcal{C}_V(t_R,t_L)$ of a two-sided CGHS black hole
is then given by the maximal volume on the set of spacelike curves $(y^+(s),y^-(s))$ with fixed endpoints at 
$t_L$ and $t_R$ on the left and right anchor curves.
This amounts to maximizing the volume functional,
\begin{equation}\label{eq:volfncl}
V=\int ds \, \sqrt{ - \dot{y}^+ \dot{y}^-(M-y^+y^-)} \,,
\end{equation}
and evaluating the resulting maximal volume. 

In order to proceed, we note that the functional is invariant under the transformations
\begin{equation}\label{eq:e_trafo}
	\begin{aligned}
		y^{+} & \mapsto e^{\epsilon}y^{+}\,,\\
		y^{-} & \mapsto e^{-\epsilon}y^-
		\,,
		\end{aligned}
\end{equation}
for $\epsilon\in\mathbb{R}$. The corresponding conserved quantity is given by
\begin{equation}\label{eq:vol_e}
	E = \sqrt{\frac{M-y^{+}y^{-}}{-\dot{y}^{+}\dot{y}^{-}}}\left(\dot{y}^{+}y^- - y^{+}\dot{y}^{-}\right).
\end{equation}
We construct the corresponding maximum volume curve by first focusing on the 
outside region on the right and rewriting the conserved charge in the $(t,\sigma)$ coordinates, 
\begin{equation}\label{eq:vol_e2}
	E = -2e^\sigma  \frac{dt}{d\sigma} \sqrt{\frac{M+e^{2\sigma}}{1-\left(\frac{dt}{d\sigma}\right)^2}}\,.
\end{equation}
This integrates to 
\begin{equation}\label{eq:tcurve}
t- t_0 = -\sigma +\frac12 \log\left(E^2+2M e^{2\sigma}+E\sqrt{E^2+4Me^{2\sigma}+4e^{4\sigma}}\right).
\end{equation}
We then convert back to Kruskal coordinates by using $y^\pm =\pm e^{\pm t +\sigma}$ and obtain the
following equation for a maximal volume curve, which is valid over the entire extended spacetime,
\begin{equation}\label{eq:ycurve}
e^{-2t_0}\big(y^+\big)^2+4My^+y^- +4e^{2t_0}\big(M^2{-}E^2\big)\big(y^-\big)^2-2E^2=0\,.
\end{equation}
Maximal volume curves that extend between the anchor curves correspond to a conserved charge in 
the range $-M<E<M$, and can be parametrised as
\begin{equation}\label{eq:parametrised}
	y^{+}=\sqrt{2} e^{t_{0}}\epsilon\sinh\tau
	\,,
	\quad
	y^{-}=-\frac{e^{-t_{0}}}{\sqrt{2}}\sinh(\tau-\mu)
	\,,
\end{equation}
with $\epsilon=\sqrt{M^2-E^2}$ and $\tanh\mu=\frac{E}{M}$, while 
curves satisfying \eqref{eq:ycurve} with $|E|>M$ run into the curvature singularity. The parameter $\tau$ 
in \eqref{eq:parametrised} runs from a negative value $\tau_L<0$ at the endpoint on the left anchor curve
to $\tau=0$, where the curve enters the black hole from the left. At $\tau=\mu$ the curve exits the black 
hole to the right and reaches the endpoint on the right anchor curve at $\tau=\tau_R$. We illustrate this 
setup in Figure \ref{fig:eternal}.

The maximal volume curve is labelled by $E$ and $t_0$ in \eqref{eq:parametrised} but these labels are
in one-to-one correspondence with the Schwarzschild times $t_L$ and $t_R$ where the curve
meets the anchor curves. 
To see this, consider the intersection points between
the maximal volume curve and the anchor curves. On the one hand we have
\begin{equation}\label{eq:parameterrelations1}
e^{2\sigma_a}=\epsilon \sinh \tau_R \sinh(\tau_R-\mu)=\epsilon \sinh(- \tau_L) \sinh(\mu-\tau_L)\,,
\end{equation}
relating curve parameters to the spatial location of the anchor curves, and on the other hand
a pair of relations involving the Schwarzschild times at the anchor points,
\begin{equation}\label{eq:parameterrelations2}
	e^{2t_R}=2 \epsilon e^{2t_0}\frac{\sinh(\tau_R)}{\sinh(\tau_R-\mu)}
	\,,
	\quad
	e^{2t_L}=\frac{e^{-2t_0}}{2\epsilon}\frac{\sinh(\mu-\tau_L)}{\sinh(-\tau_L)}
	\,.
\end{equation}
The second equation in \eqref{eq:parameterrelations1} is satisfied by imposing $\tau_R=\mu-\tau_L$ and 
the time relations can then be re-expressed as
\begin{equation}\label{eq:timerelations}
e^{t_R-t_L}=2\epsilon e^{2t_0}
	\,,
	\quad
e^{t_R+t_L}=\frac{\sinh(\tau_R)}{\sinh(\tau_R-\mu)}
	\,.
\end{equation}
By combining \eqref{eq:parameterrelations1} with the second equation in \eqref{eq:timerelations}
and doing some algebra one eventually arrives at
\begin{equation}\label{eq:epsilonrelation}
\epsilon=-e^{2\sigma_a} \cosh(t_R{+}t_L)+\sqrt{e^{4\sigma_a} \cosh^2(t_R{+}t_L)+2Me^{2\sigma_a}+M^2}\,.
\end{equation}
The parametrisation \eqref{eq:parametrised} can then be re-expressed in terms of $t_R$ and $t_L$, as
\begin{equation}\label{eq:reexpressed}
	y^{+}=e^{\frac12 (t_R-t_L)}\sqrt{\epsilon}\sinh\tau
	\,,
	\quad
	y^{-}=-e^{-\frac12 (t_R-t_L)}\sqrt{\epsilon}\sinh(\tau-\mu)
	\,.
\end{equation}
with $\epsilon(t_R{+}t_L)$ given by \eqref{eq:epsilonrelation}.

The volume functional \eqref{eq:volfncl} is easily evaluated in this parametrisation, 
\begin{equation}\label{eq:2sidedvolume}
	\begin{aligned}
		V & = \epsilon \int_{\tau_L}^{\tau_R}  d\tau \,\cosh{\tau}\cosh{(\tau-\mu)} \\
		& =\frac{M}{2}\big(2\tau_R-\mu\big) +\frac{\epsilon}{2}\sinh{(2\tau_R-\mu)}
		\,.
		\end{aligned}
\end{equation}
The first equation in \eqref{eq:parameterrelations1} can be combined with second equation in \eqref{eq:timerelations}
to give 
\begin{equation}\label{eq:2tauminusmu}
2\tau_R-\mu= \textrm{arccosh}{\left[\frac{1}{M}\left(e^{2\sigma_a} \cosh(t_R{+}t_L)
{+}\sqrt{e^{4\sigma_a} \cosh^2(t_R{+}t_L){+}2Me^{2\sigma_a}{+}M^2}\,\right)\right]},
\end{equation}
which can then be inserted in \eqref{eq:2sidedvolume} to obtain an exact, if somewhat unwieldy, 
formula for the maximal volume as a function of $t_R{+}t_L$.

The expression for the volume simplifies enormously at late times,
\begin{equation}\label{eq:2latetimeV}
V=\frac{M}{2} (t_R{+}t_L)+e^{2\sigma_a}+\frac{M}{2}\left(1+2\sigma_a-\log{\frac{M}{2}}\right)
+\mathcal{O} \big(e^{-2(t_R{+}t_L)}\big)\,,
\end{equation}
with a leading term that grows linearly with time, followed by a constant term that depends on 
the location of the anchor curve, and subsequent terms that are exponentially suppressed at
late times. As expected, the volume diverges in the $\sigma_a\to\infty$ limit, where the anchor 
curves are moved off to spatial infinity, but the late time rate of growth is unaffected by the 
location of the anchor curves. The volume prescription for complexity is sometimes taken to 
only include the volume inside the event horizon of the black hole. This amounts to cutting 
off the volume integration in \eqref{eq:2sidedvolume} at the event horizon,
\begin{equation}\label{eq:Veh}
\begin{aligned}
V_{\text{EH}} & =\epsilon \int_{0}^{\mu}  d\tau \,\cosh{\tau}\cosh{(\tau-\mu)} \\
                      & =\frac{M}{2}\big(\mu+\tanh\mu\big)
	              \,,
\end{aligned}
\end{equation}
which can be shown to grow at the same rate at late times as the full volume between 
anchor curves. Later on, when we consider dynamical black holes formed by 
the gravitational collapse of matter, we will see that the stretched horizon is a natural choice of 
anchor curve. In the case at hand, we define the stretched horizon to be a membrane outside the 
black hole, with an area that is one unit larger than the area of the event horizon,
\begin{equation}\label{eq:SHdef}
e^{-2\phi_{\text{SH}}}=e^{-2\phi_{\text{EH}}}+1=M+1
	\,.
\end{equation}
This is a curve of constant dilaton field outside the black hole, which is how we defined our anchor curves
above. Indeed, with this definition, the stretched horizon is located at $\sigma_\text{SH}=0$ in the $(t,\sigma)$
coordinate system and the volume inside the stretched horizon can be obtained by 
setting $\sigma_a=0$ in \eqref{eq:2latetimeV}.
The result differs from the volume inside the event horizon by only a small amount 
and the late time volume growth is the same. For a static two-sided black hole the 
location of the anchor curve is unimportant if all we are interested in is the late time
rate of growth of the complexity. It is only when we consider dynamical 
black holes that the advantage of using the stretched horizon as the anchor curve 
becomes apparent.

We note that the Hawking temperature of a CGHS black hole is $T_\text{H}=\frac{1}{2\pi}$, independent
of the black hole mass \cite{Callan:1992rs}, and the Bekenstein-Hawking entropy is given by
$S_\text{BH}=2 M$. The late time rate of growth of the complexity given by the volume in \eqref{eq:2latetimeV}
is thus proportional to $S_\text{BH}\,T_\text{H}$, which is precisely in line with the original $\mathcal{C} = V$ 
proposal \cite{Susskind:2014rva}.
\subsection{Complexity of a black hole formed by gravitational collapse}
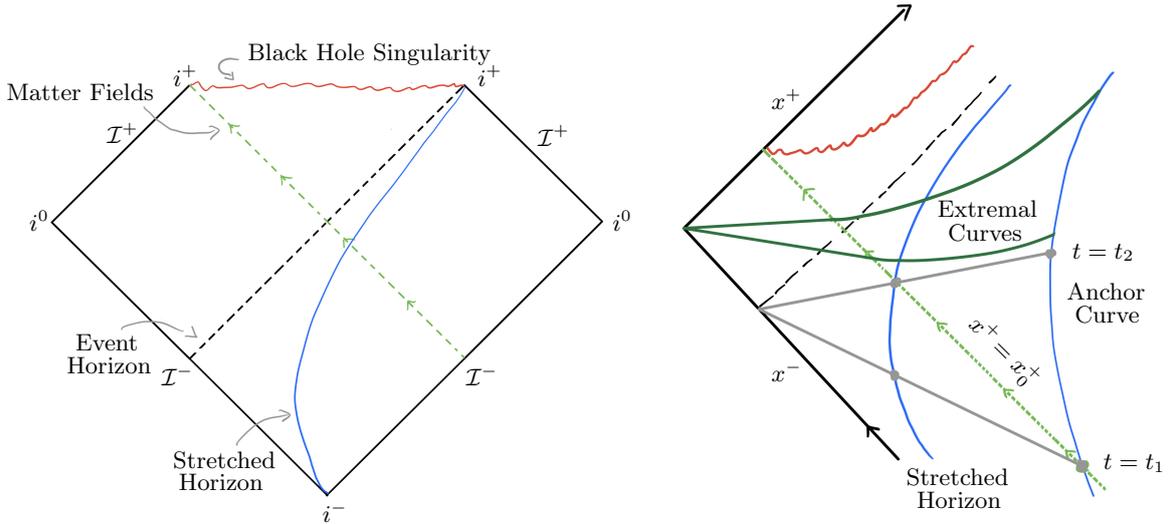
\begin{figure}[h]
	\begin{center}
		\begin{overpic}[width=.50\textwidth]{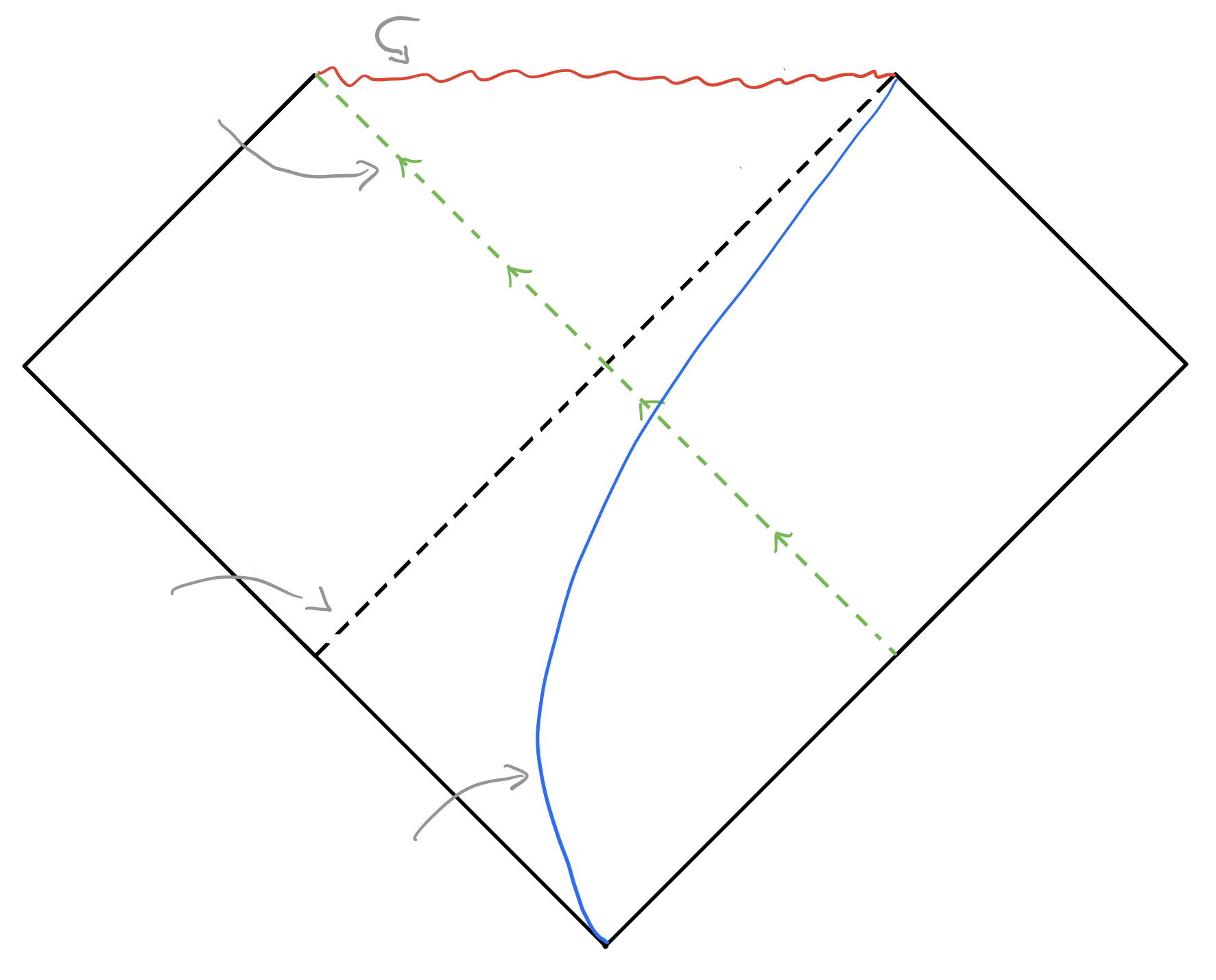}
			\put (-6,72) {\footnotesize{Matter Fields}}
			\put (36,79) {\footnotesize{Black Hole Singularity}}
			\put (23,7.5) {\footnotesize{Stretched}}
			\put (24,3.5) {\footnotesize{Horizon}}
			\put (6,28) {\footnotesize{Event}}
			\put (4.5,24) {\footnotesize{Horizon}}
			\put (21,22) {\footnotesize{$\mathcal{I}^{-}$}}
			\put (74,22) {\footnotesize{$\mathcal{I}^{-}$}}
			\put (86.5,64) {\footnotesize{$\mathcal{I}^{+}$}}
			\put (11.5,64.5) {\footnotesize{$\mathcal{I}^{+}$}}
			\put (-2,49) {\footnotesize{$i^{0}$}}
			\put (99.5,49) {\footnotesize{$i^{0}$}}
			\put (49,-2) {\footnotesize{$i^{-}$}}
			\put (23,74.5) {\footnotesize{$i^{+}$}}
			\put (76,74.5) {\footnotesize{$i^{+}$}}
		\end{overpic}
		\hspace{.5cm}
		\begin{overpic}[width=.39\textwidth]{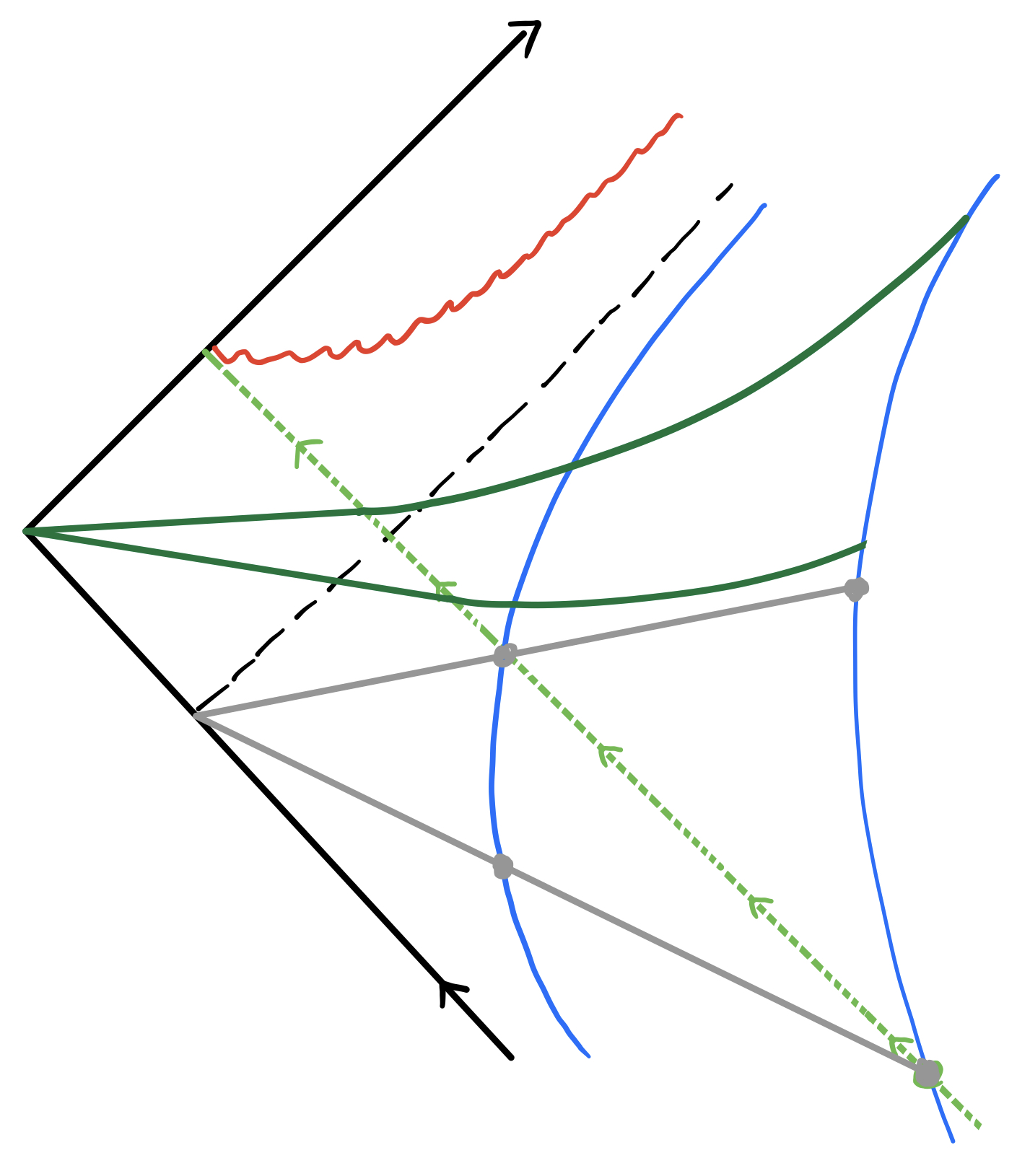}
			\put (19,25) {\footnotesize{$x^{-}$}}
			\put (19,78) {\footnotesize{$x^{+}$}}
			\put (56,36) {\rotatebox{-45}{\footnotesize{$x^{+}\!=x^{+}_{0}$}}}
			\put (45,5) {\footnotesize{Stretched}}
			\put (47,.5) {\footnotesize{Horizon}}
			\put (76,41) {\footnotesize{Anchor}}
			\put (77.5,36.5) {\footnotesize{Curve}}
			\put (77,49) {\footnotesize{$t=t_{2}$}}
			\put (83,8) {\footnotesize{$t=t_{1}$}}
			\put (51,57) {\footnotesize{Extremal}}			
			\put (53,52.5) {\footnotesize{Curves}}
		\end{overpic}
	\end{center}
\vspace{-0.3cm}
\caption{(\textit{Left})
This cartoon depicts the Penrose diagram of a one-sided CGHS black hole formed by gravitational collapse. (\textit{Right}) This is a Kruskal diagram of the Penrose diagram on the left. The color coding coincides. The gray line is an equal time curve.}\label{fig:collapse_1}
\end{figure}

Next, we consider collapsing a thin shell of matter at $x^{+}=x^{+}_{0}$, mediated by matter fields $f_{i}$, into a CGHS vacuum and creating a black hole. This amounts to
\begin{equation}
	T^{f}_{++}
	=
	\frac{M}{x_{0}^{+}}\delta(x^{+}-x^{+}_{0})
	\,,
\end{equation}
where $\delta(x^{+}-x^{+}_{0})$ is a delta function. For the geometry this implies
\begin{equation}
	e^{-2\phi}
	=
	e^{-2\rho}
	=
	\begin{cases}
		-x^{+}x^{-}& \text{if } x^{+}< x^{+}_{0}\,,\\
    		-x^{+}\left(x^{-}+\frac{M}{x_{0}^{+}}\right)+M            &  \text{if } x^{+}\geq x^{+}_{0}\,.
	\end{cases}
\end{equation}
The resulting black hole is one-sided, as illustrated in Figure \ref{fig:collapse_1}.

Just as in the eternal black hole geometry, we need to introduce an anchor curve outside the black hole to obtain a finite volume. The extremal curve now reaches from a point on the anchor curve to a point on $x^{-}x^{+}=0$. In Figure \ref{fig:collapse_1}, we sketch the setup.
In contrast to the eternal black hole where we had two anchor points and therefore a unique extremal curve connecting these two points, there is now only one anchor point and therefore we have to supply a prescription for additional boundary conditions. In~\cite{Chapman:2018dem}, for instance, it was argued that in order to obtain a smooth volume at the radial origin, the 
additional boundary conditions should be $t'(r)=0$ at $r=0$. Expressing the corresponding condition in our setup in Kruskal coordinates one finds the relation
\begin{equation}\label{eq:kruskal_bc}
	x^{-} \dot{x}^{+} -x^{+} \dot{x}^{-}=0
	\,,
\end{equation}
at $x^{+} x^{-} = 0$. 
An alternative prescription is to consider all locally extremal curves originating from the anchor point, and selecting the curve that maximizes the volume inside the black hole. This computation can be done and, interestingly, it turns out that this prescription leads to exactly the
same curves as the boundary conditions~\eqref{eq:kruskal_bc}.

Curves that maximise the volume functional \eqref{eq:volume_prop_mod}  in the one-sided black hole background 
are obtained by patching together maximal curves
across the infalling shockwave. In the inside region, $0<x^+<x^+_0$, the geometry is flat and the volume functional
takes the simple form,
\begin{equation}\label{eq:simplevolume}
V=\int ds \, \sqrt{ y^+\dot{y}^+ y^-\dot{y}^-} \,.
\end{equation}
One obtains a two-parameter family of maximal curves,
\begin{equation}\label{eq:maxcurves1}
\big(x^+\big)^2=\alpha \big(x^-\big)^2+\beta
\end{equation}
where $\alpha>0$ and $\beta<\sqrt{x^{-}_0}$ are real valued parameters. The boundary condition \eqref{eq:kruskal_bc}
selects curves with $\beta=0$, which are simply straight lines emanating from the origin $x^+=x^-=0$ in Kruskal coordinates
(see Figure~\ref{fig:collapse_1}).

In the outside region, $x^+>x^+_0$, the geometry is that of a static black hole and the volume functional reduces to 
\eqref{eq:volfncl} in shifted Kruskal coordinates,
\begin{equation}
y^+=x^+\,,\qquad y^-=x^-+\frac{M}{x_0^+}\,.
\end{equation}

There is again a two-parameter family of maximal volume curves satisfying \eqref{eq:ycurve} and labelled by $E$ and $t_{0}$. We use Weierstra\ss-Erdmann conditions to patch across the shockwave. First of all, the curve itself should be continuous. A second condition comes from viewing the integral in the volume functional as a Lagrange density and requiring that the momenta conjugate to $y^{+}$ and $y^{-}$ be continuous across the shock. Those matching conditions uniquely determine the parameter $\alpha$ and $\beta$ in terms of $E$ and $t_{0}$, and vice versa. One finds, in particular, that curves with $\beta=0$ in the inside region match onto curves with $E=M$ on the outside leaving us with a one-parameter family
\begin{equation}\label{eq:curve_para}
	y^{-} = 
	\begin{cases}
		\frac{M}{2y^{+}}-\frac{e^{-2t_{0}}}{4M}y^{+} & y^{+}>x_{0}^{+} \,,\\
		\frac{M}{x_{0}^{+}}-\left(\frac{M}{2x^{+}_{0}}+\frac{e^{-2t_{0}}}{4M}\right)y^{+} & y^{+}<x_{0}^{+} \,.
	\end{cases}
\end{equation}
Subsequently, the remaining parameter $t_{0}$ can be uniquely related to the tortoise time $t$ on the anchor curve, see Figure~\eqref{fig:collapse_1}. As in section~\ref{subsec:eternal}, we define the anchor curves, parametrised by $\sigma_a$, so that the dilaton field is constant $\phi=\phi_a$,
\begin{equation}\label{eq:anch_def}
	e^{2\sigma_a} = e^{-2\phi_a}-M = 
	\begin{cases}
		-y^{+}y^{-} & y^{+}>x_{0}^{+}\\
		-y^{+}\left(y^{-}-\frac{M}{x_{0}^{+}}\right) - M & y^{+}<x_{0}^{+}
		\end{cases}
	\, .
\end{equation}
We now obtain a relation between $\sigma_a$, $t_0$ and $t$ by combining~\eqref{eq:curve_para} and \eqref{eq:anch_def}.
One is now in a position to evaluate the volume functional~\eqref{eq:volume_prop_mod}.

The growth of the volume inside the stretched horizon as a function of tortoise time $t$ is given by
\begin{equation}
	V_{\text{SH}}^{\prime}(t)
	=
		\begin{cases}
			\frac{1}{2}M & t \geq t_{2}\,,\\
			\frac{1}{2}M\cosh(t)^{-2} + \mathcal{O}(\frac{1}{M}) & t < t_{2}\,,
		\end{cases}
\end{equation}
where $t_{2}$ is the moment the black hole is formed. We conclude that the volume growth sets in essentially at the time the black hole is formed, which is consistent with causality.

\begin{figure}[h]
		\begin{centering}
		\begin{overpic}[width=.7\textwidth]{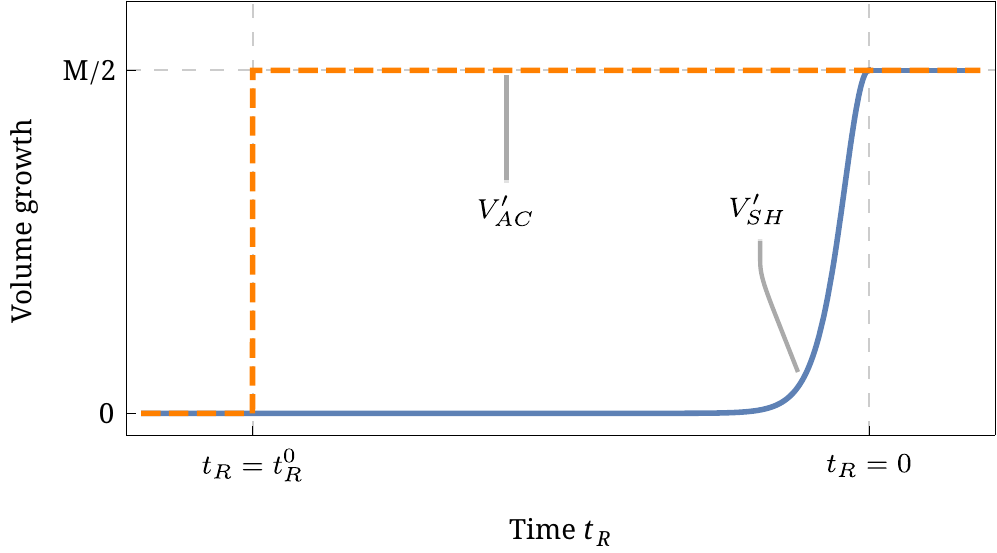}
		\put(20,-0.3){\includegraphics[scale=0.5]{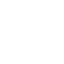}}
		\put(22,8){\footnotesize{$t=t_{1}$}}
		\put(50,-0.3){\includegraphics[scale=0.55]{figures/patch.png}}
		\put(51,1){\footnotesize{Time $t$}}
		\put(83,-0.3){\includegraphics[scale=0.55]{figures/patch.png}}
		\put(84,8){\footnotesize{$t=t_{2}$}}
		\end{overpic}
		\par
		\end{centering}
	\caption{This graph demonstrates the difference between the volume growth $V'_{SH}$ and $V'_{AC}$. $V'_{AC}$ starts growing at the time the shockwave crosses the anchor curve $t_{1}$ which is much earlier than the time $t_{2}$ the black hole is created.}
	\label{fig:cv}
\end{figure}

In contrast to the aforementioned, we could consider the total volume up to an arbitrary anchor line.\footnote{In particular, we could imagine this anchor line to be asymptotically far away, in analogy to computations done in the AdS/CFT setup.} The result agrees at times $t>t_{2}$ but disagrees strongly before that time. We have
\begin{equation}
	V_{\text{AC}}^{\prime}(t)=
		\begin{cases}
			\frac{1}{2}M & t \geq t_{1}\,,\\
			0 & t < t_{1}\,,
		\end{cases}
		\,
\end{equation}
where $t_{1}$ is the time when the anchor curve crosses the shockwave line, see Figure \ref{fig:collapse_1}. This is long before the black hole is created\footnote{In fact, as we move the anchor curve infinitely far away,  the time difference of black hole creation and $t_R^1$ also goes to infinity.} and it would imply that complexity starts growing at the moment the shockwave is released, see Figure~\ref{fig:cv}. 
For this reason, we prefer to use the stretched horizon as the anchor curve and only 
consider the volume inside the black hole. The volume integral can be cut off at either 
the event horizon or at the stretched horizon.

We can compare the complexity growth of the gravitational collapse model to the complexity growth of the eternal black hole, both at late time. The result for the gravitational collapse model is $V'=M/2$ for $t>t_{2}$, which is exactly half of the eternal black hole result, see \eqref{eq:Veh}. This factor $\frac{1}{2}$ corresponds to the fact that in the current case we only consider half of the volume slice (since we only have a one-sided black hole), as compared to the eternal black hole case (which is two-sided).
\section{The RST model: Complexity in a semi-classical black hole}
The CGHS model can be extended such that one can study a semi-classical black hole 
analytically~\cite{Bilal:1992kv,deAlwis:1992emy}. Here we adopt a particular modification due
to RST~\cite{Russo:1992ht}, which is given by
\begin{equation}\label{eq:rst_model}
	S_{\text{RST}}  = S_{\text{CGHS}} + S_{\text{q}} + S_{\text{ct}}
	\,.
\end{equation}
If one takes $N$, the number of matter fields $f_{i}$, to be large, then $S_{\text{q}}$ represents the leading order 
quantum correction due to matter, which in conformal gauge reads
\begin{equation}
	S_{q}
	=
	-\kappa\int_{\mathcal{M}} d^{2} x \, \partial_{+}\rho\partial_{-}\rho
	\,.
\end{equation}
Here $\kappa:=N/12$ can be thought of as playing the role of $\hbar$, which is put to unity. 
The additional RST counter term has the following form in conformal gauge,
\begin{equation}
	S_{ct}
	=
	-\kappa \int_{\mathcal{M}} d^{2}x \, \phi \partial_{+}\partial_{-}\rho
	\,.
\end{equation}
This term is allowed by the symmetries of the model and when it is added the semi-classical 
field equations take a particularly simple form and are easily solved analytically.

The solutions of the equations of motion can be written in compact form if one defines 
new field variables,
\begin{equation}\label{eq:rstfields}
	\sqrt{\kappa}\Omega
	:=
		e^{-2\phi}+\frac{\kappa}{2}\phi
		\,,
		\quad
	\sqrt{\kappa}\chi
	:=
	e^{-2\phi}
	-
	\frac{\kappa}{2}\phi
	+
	\kappa \rho
	\,.
\end{equation}
Using the new fields, the semi-classical action reads
\begin{equation}
	S_{\text{RST}}
	=
	2\int_{\mathcal{M}}d^{2}x\left[
		-\partial_{+}\chi\partial_{-}\chi
		+
		\partial_{+}\Omega
		\partial_{-}\Omega
		+
		e^{\frac{2}{\sqrt{\kappa}}(\chi-\Omega)}
		+
		\frac{1}{2}\sum^{N}_{i=1}\partial_{+}f_{i}\partial_{-}f_{i}
	\right]
	\,.
\end{equation}
The RST model continues to enjoy the symmetry of the classical theory that allowed us to choose 
Kruskal coordinates, setting $\phi=\rho$ and as a result $\Omega=\chi$.

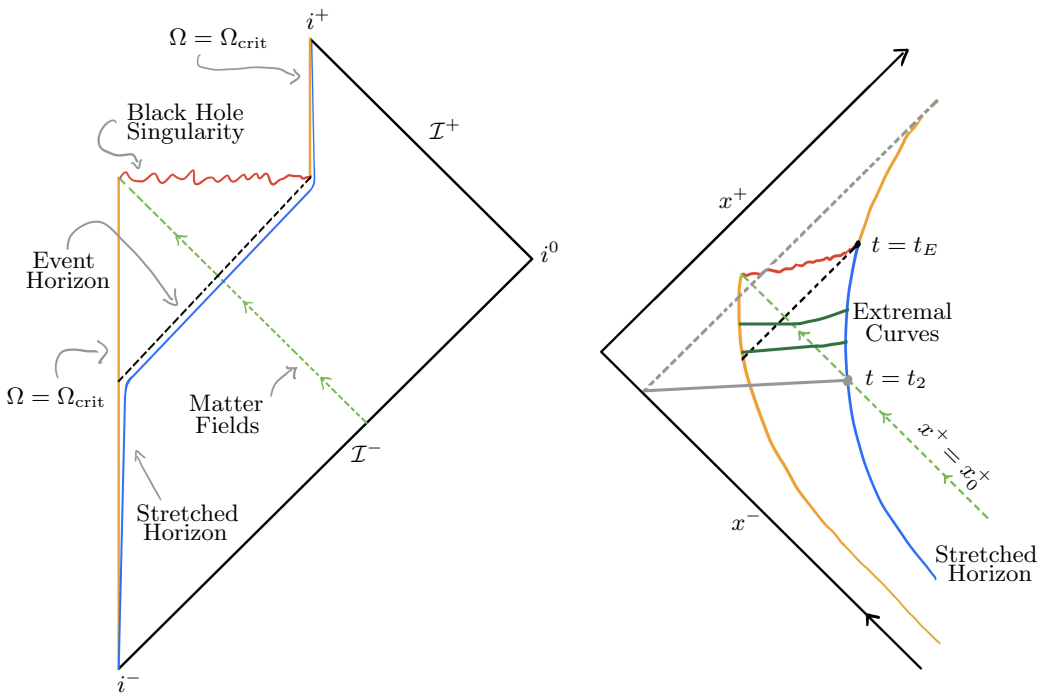
\begin{figure}[h]
	\begin{center}
		\begin{overpic}[width=.43\textwidth]{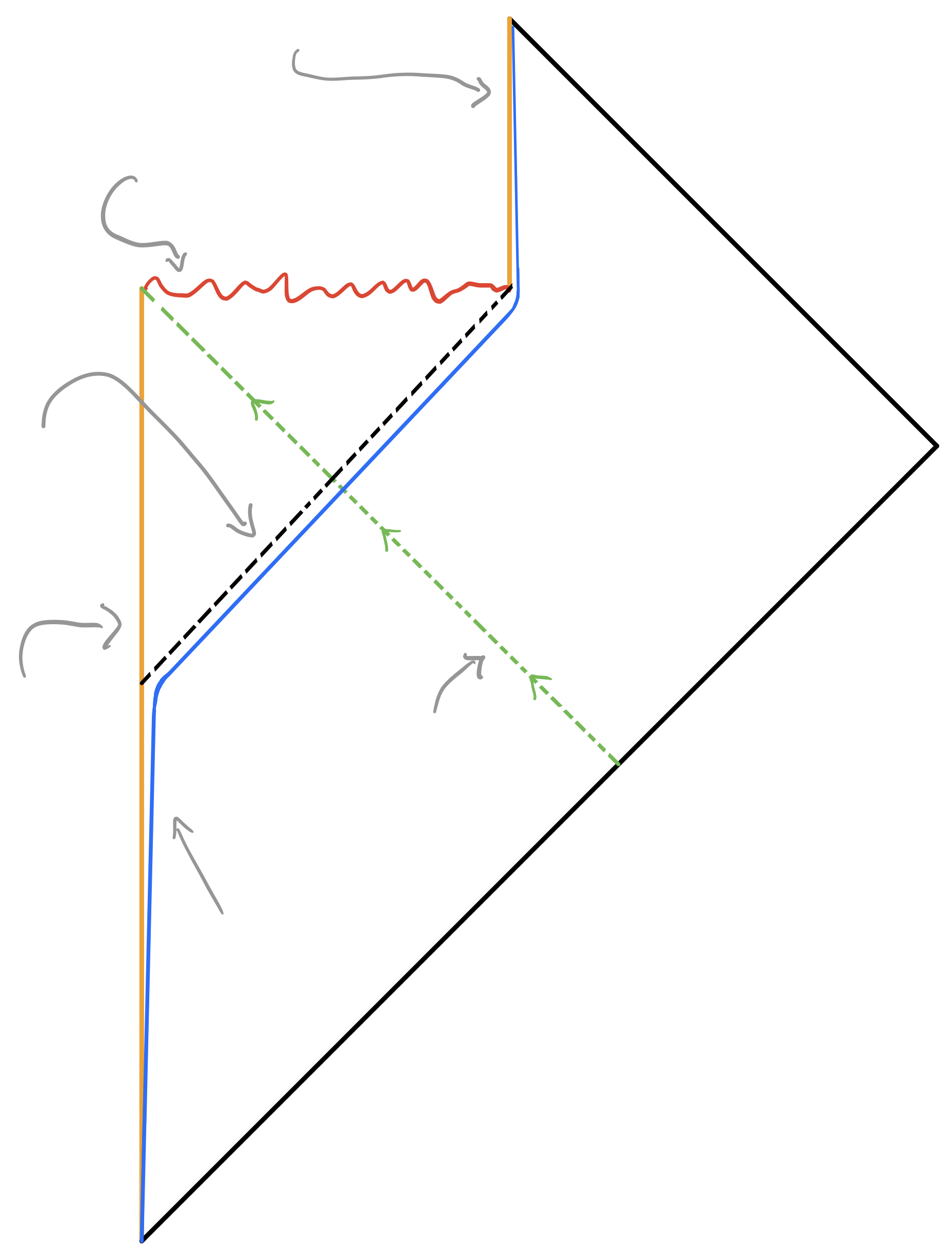}
			\put (-6,42) {\footnotesize{$\Omega=\Omega_{\text{crit}}$}}
			\put (19,97.5) {\footnotesize{$\Omega=\Omega_{\text{crit}}$}}
			\put (14,24) {\footnotesize{Stretched}}
			\put (15,21) {\footnotesize{Horizon}}
			\put (12.5,86) {\footnotesize{Black Hole}}
			\put (12.5,83) {\footnotesize{Singularity}}
			\put (-2,63) {\footnotesize{Event}}
			\put (-3,60) {\footnotesize{Horizon}}
			\put (22,41) {\footnotesize{Matter}}
			\put (23,37.5) {\footnotesize{Fields}}
			\put (40,100) {\footnotesize{$i^{+}$}}
			\put (76,64) {\footnotesize{$i^{0}$}}
			\put (11,-2.5) {\footnotesize{$i^{-}$}}
			\put (47,33) {\footnotesize{$\mathcal{I}^{-}$}}
			\put (59,83) {\footnotesize{$\mathcal{I}^{+}$}}
		\end{overpic}
		\hspace{.5cm}
		\begin{overpic}[width=.35\textwidth]{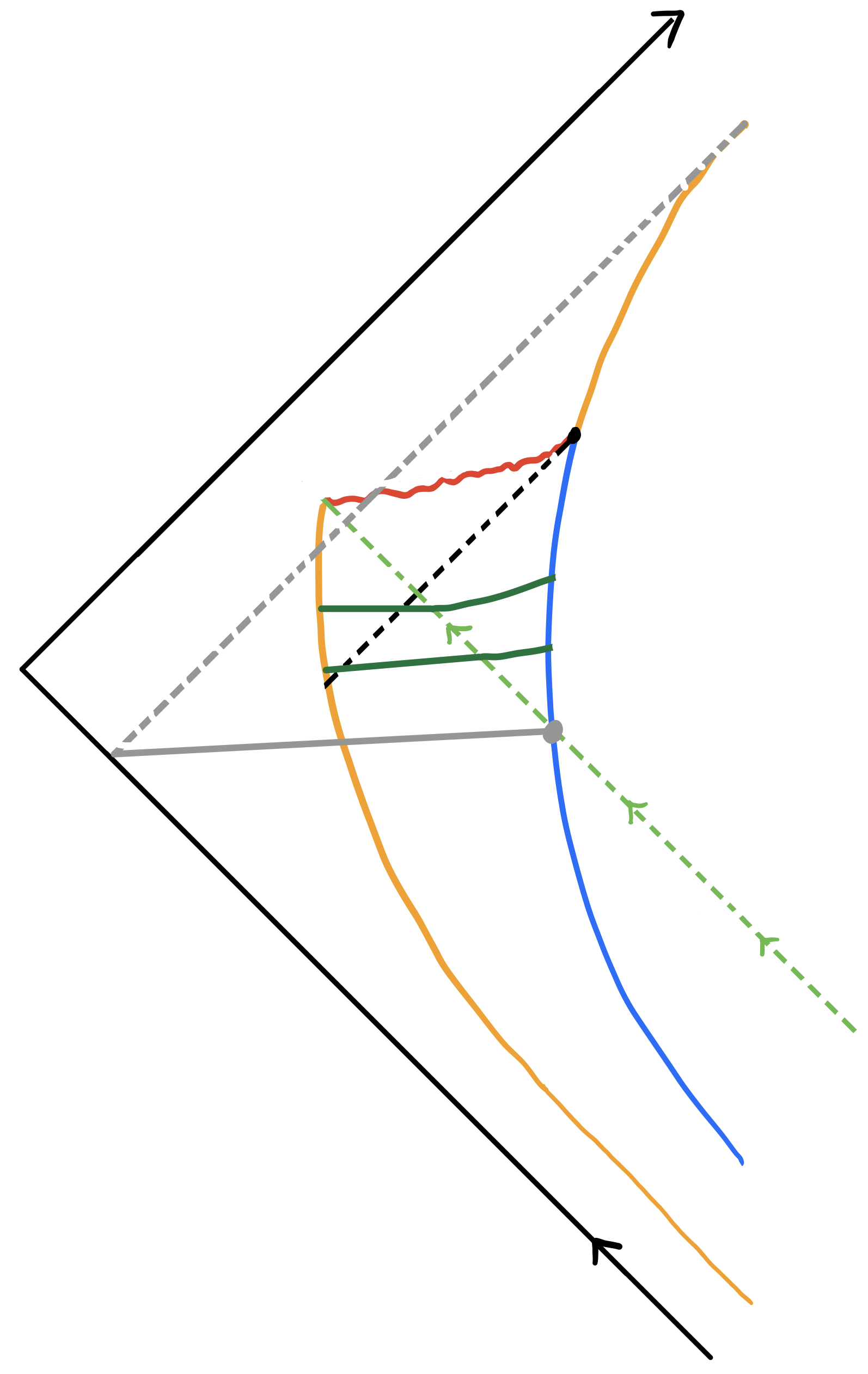}
			\put (22,23) {\footnotesize{$x^{-}$}}
			\put (20,74) {\footnotesize{$x^{+}$}}
			\put (56,15) {\footnotesize{Horizon}}
			\put (54,18) {\footnotesize{Stretched}}
			\put (50.5,39) {\rotatebox{-45}{\footnotesize{$x^{+}\!=x^{+}_{0}$}}}
			\put (41,56.5) {\footnotesize{Extremal}}			
			\put (43,53) {\footnotesize{Curves}}
			\put (43,46) {\footnotesize{$t=t_{2}$}}
			\put (44,67) {\footnotesize{$t=t_{E}$}}
		\end{overpic}
	\end{center}
\vspace{-0.5cm}
\caption{
(\textit{Left}) Cartoon of the Penrose diagram of the life cycle of an evaporating black hole formed by collapse. (\textit{Right}) Depicts Kruskal diagram of the scenario illustrated in the left figure. The color coding coincides.
}\label{fig:evaporation}
\end{figure}

Again, we study an incoming shockwave of energy $M$ at $x^{+}=x^{+}_{0}$ of the form
\begin{equation}\label{eq:qmcollapse}
	T_{++}^f
	=
	\frac{M}{x^{+}_{0}}
	\delta(x^{+}-x^{+}_{0})
	\,.
\end{equation}
The semi-classical collapse solution of interest, given in Kruskal coordinates, is
\begin{equation}\label{eq:q_coll_sol_1}
	\sqrt{\kappa}\Omega(x^{+},x^{-}) =
		-x^{+}x^{-}+\left(x^{+}_{0}-x^{+}\right)\frac{M}{x^{+}_0}\Theta(x^{+}-x^{+}_{0})
		-\frac{\kappa}{4}\ln\left(-x^{+}x^{-}\right)
	\,.
\end{equation}
The solution describes flat spacetime for $x^{+} \leq x^{+}_0$ and an evaporating black hole for $x^{+} > x^{+}_0$,
as shown in Figure~\ref{fig:evaporation}.
It is important to note that in the RST model only those regions of spacetime where 
$\Omega(x^+,x^-) \geq \Omega_\text{crit} := \frac{\sqrt{\kappa}}{4} \left(1 - \log \left(\frac{\kappa }{4}\right)\right)$ are 
considered physical. 
In the flat spacetime region inside the infalling shell, the boundary of the physical region is a timelike curve that can be interpreted as the origin in spherical coordinates in a higher-dimensional parent theory. 
For $x^+>x^+_0$, the curve $(x^{+}_S,x^{-}_S)$ for which $\Omega(x^{+}_S,x^{-}_S)=\Omega_\text{crit}$ turns 
spacelike and defines the location of the black hole singularity. 
The semi-classical black hole evaporates and eventually the singularity terminates at an endpoint, after which
the solution can be extended into a late time flat region where the physical boundary is again timelike. A more
detailed description of the semi-classical geometry can for instance be found in \cite{Thorlacius:1994ip}.

Although we have closed expressions for the solutions of the RST model, we were not able to solve the extremization problem of a spacelike volume analytically and in order to check whether the complexity as volume prescription gives results consistent with general expectations we had to resort to numerical methods.

\begin{figure}[h]\center
	\subfloat[$M/\kappa=1$\label{fig:cv_numeric_plots_a}]{
		\begin{centering}
		\begin{overpic}{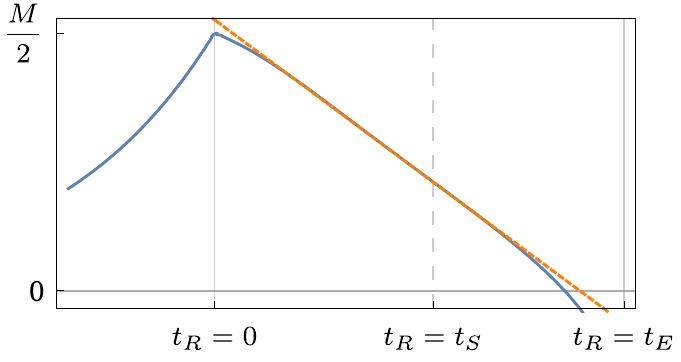}							\put (-5,28) {\footnotesize{$V'(t)$}}
			\put(83,-2.9){\includegraphics[scale=0.25]{figures/patch.png}}
			\put(91,-2.9){\includegraphics[scale=0.25]{figures/patch.png}}
			\put(25,-2.9){\includegraphics[scale=0.25]{figures/patch.png}}
			\put(33,-2.9){\includegraphics[scale=0.25]{figures/patch.png}}
			\put(55,-2.9){\includegraphics[scale=0.25]{figures/patch.png}}
			\put(63,-2.9){\includegraphics[scale=0.25]{figures/patch.png}}
			\put(27,2){\footnotesize{$t=t_{2}$}}
			\put(59,2){\footnotesize{$t=t_{S}$}}
			\put(87,2){\footnotesize{$t=t_{E}$}}
		\end{overpic}
		\par\end{centering}
	}\hfill
	\subfloat[$M/\kappa=5$]{
		\begin{centering}
		\begin{overpic}{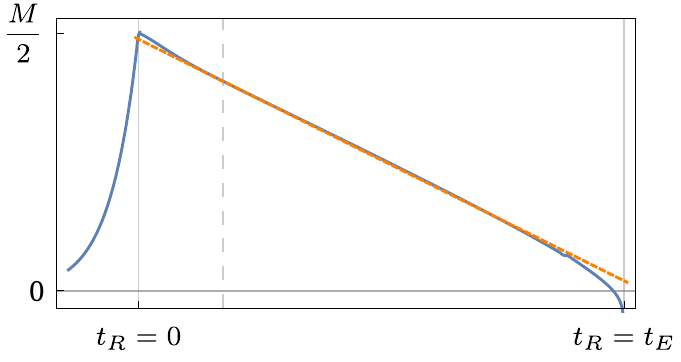}
			\put (-5,28) {\footnotesize{$V'(t)$}}
			\put(14,-2.9){\includegraphics[scale=0.25]{figures/patch.png}}
			\put(17,-2.9){\includegraphics[scale=0.25]{figures/patch.png}}
			\put(83,-2.9){\includegraphics[scale=0.25]{figures/patch.png}}
			\put(91,-2.9){\includegraphics[scale=0.25]{figures/patch.png}}
			\put(25,-2.9){\includegraphics[scale=0.25]{figures/patch.png}}
			\put(33,-2.9){\includegraphics[scale=0.25]{figures/patch.png}}
			\put(55,-2.9){\includegraphics[scale=0.25]{figures/patch.png}}
			\put(63,-2.9){\includegraphics[scale=0.25]{figures/patch.png}}
			\put(16,2){\footnotesize{$t=t_{2}$}}
			\put(34,20){\footnotesize{$\leftarrow t=t_{S}$}}
			\put(87,2){\footnotesize{$t=t_{E}$}}
		\end{overpic}
		\par\end{centering}
	}\hfill

	\subfloat[$M/\kappa=50$\label{fig:cv_numeric_plots_large_m}]{
		\begin{centering}
		\begin{overpic}{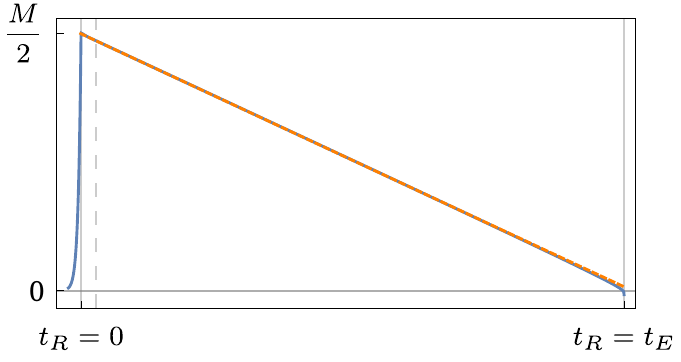}
			\put (-5,28) {\footnotesize{$V'(t)$}}
			\put(2,-2.9){\includegraphics[scale=0.25]{figures/patch.png}}
			\put(6,-2.9){\includegraphics[scale=0.25]{figures/patch.png}}
			\put(10,-2.9){\includegraphics[scale=0.25]{figures/patch.png}}
			\put(14,-2.9){\includegraphics[scale=0.25]{figures/patch.png}}
			\put(17,-2.9){\includegraphics[scale=0.25]{figures/patch.png}}
			\put(83,-2.9){\includegraphics[scale=0.25]{figures/patch.png}}
			\put(91,-2.9){\includegraphics[scale=0.25]{figures/patch.png}}
			\put(25,-2.9){\includegraphics[scale=0.25]{figures/patch.png}}
			\put(33,-2.9){\includegraphics[scale=0.25]{figures/patch.png}}
			\put(55,-2.9){\includegraphics[scale=0.25]{figures/patch.png}}
			\put(63,-2.9){\includegraphics[scale=0.25]{figures/patch.png}}
			\put(7,2){\footnotesize{$t=t_{2}$}}
			\put(16,20){\footnotesize{$\leftarrow t=t_{S}$}}
			\put(87,2){\footnotesize{$t=t_{E}$}}
		\end{overpic}
		\par\end{centering}
	}\hfill
	\subfloat[$M/\kappa=100$\label{fig:cv_numeric_plots_large_m_2}]{
		\begin{centering}
		\begin{overpic}{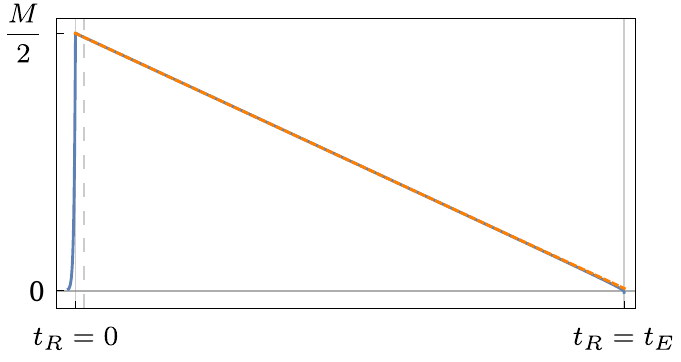}
			\put (-5,28) {\footnotesize{$V'(t)$}}
			\put(2,-2.9){\includegraphics[scale=0.25]{figures/patch.png}}
			\put(6,-2.9){\includegraphics[scale=0.25]{figures/patch.png}}
			\put(10,-2.9){\includegraphics[scale=0.25]{figures/patch.png}}
			\put(14,-2.9){\includegraphics[scale=0.25]{figures/patch.png}}
			\put(17,-2.9){\includegraphics[scale=0.25]{figures/patch.png}}
			\put(83,-2.9){\includegraphics[scale=0.25]{figures/patch.png}}
			\put(91,-2.9){\includegraphics[scale=0.25]{figures/patch.png}}
			\put(25,-2.9){\includegraphics[scale=0.25]{figures/patch.png}}
			\put(33,-2.9){\includegraphics[scale=0.25]{figures/patch.png}}
			\put(55,-2.9){\includegraphics[scale=0.25]{figures/patch.png}}
			\put(63,-2.9){\includegraphics[scale=0.25]{figures/patch.png}}
			\put(7,2){\footnotesize{$t=t_{2}$}}
			\put(14,20){\footnotesize{$\leftarrow t=t_{S}$}}
			\put(87,2){\footnotesize{$t=t_{E}$}}
		\end{overpic}
		\par\end{centering}
	}\hfill
	\caption{Numerical results of volume growth for different values of $M/\kappa$. The black hole creation time is indicated by~$t=t_{2}$ while the evaporation process is completed at time~$t=t_E$. The blue curve depicts the numerical result while the dashed orange line is obtained by a linear extrapolation of the curve around the scrambling time~$t_S$.\label{fig:cv_numeric_plots}}
\end{figure}

Since the extremization of
\begin{equation}\label{eq:volume_prop_mod_rst}
	V = \int ds \, e^{-2\phi} \sqrt{ g_{\mu \nu} \dot{y}^{\mu} \dot{y}^\nu }
\end{equation}
can be performed by solving corresponding Euler-Lagrange equations, the numerical problem is simply to solve 
a non-linear ordinary differential equation with appropriate boundary conditions, whose solutions provide a 
parametrisation of the extremized volume.
Numerically integrating \eqref{eq:volume_prop_mod_rst} inside the stretched horizon 
provides us with the volume complexity $V(t)$ at a value of tortoise time $t$ 
that is determined by the choice of anchor curve in the same way as for a classical 
dynamical black hole. If we use a curve of constant $\Omega$ far from the black hole, 
then the complexity begins to grow very early, long before the incoming shockwave 
reaches the stretched horizon. If we instead use the stretched horizon as our
anchor curve, then the complexity growth turns on essentially when the black hole is 
formed and this prescription is used in obtaining the numerical results presented in 
Figure~\ref{fig:cv_numeric_plots}. We note, however, that the choice of anchor curve 
only affects the onset time and the early growth rate of the complexity, but not the 
slope of the curve (at leading order for large $M/\kappa$) showing the decreasing 
growth rate after a scrambling time has passed from the onset of complexity growth.

Note that, like in the case of classical collapse, it is not a priori clear what the appropriate boundary conditions at the origin are. We have chosen to apply boundary conditions analogous to~\eqref{eq:kruskal_bc} at $\Omega=\Omega_{\text{crit}}$ and then use a shooting
algorithm to obtain the corresponding maximal volume curve. We apply the 
Weierstra\ss-Erdmann matching conditions to patch across the shockwave and then continue the numerical evaluation outwards.

Further, it could be reasoned that the factor $e^{-2\phi}$ in \eqref{eq:volume_prop_mod_rst}, which is interpreted as the area of the transverse two-sphere in the higher-dimensional theory, should be replaced by the quantum corrected area $\Omega - \Omega_{\text{crit}}$. We find that the slope of $V'(t)$ is not particularly sensitive to this replacement, at least not in the parameter range where our
numerical evaluation is reliable (see below).

Results for the functions $V'(t)$ for different values of $M/\kappa$ are plotted in Figure~\ref{fig:cv_numeric_plots}. We observe that for reasonably high values of $M/\kappa$ the numerical result is consistent with a linear decrease after the scrambling time~$t_{S}=\log(4 M/\kappa)$. We do not expect that the volume growth follows the linear trend forever, since eventually, the black hole is small enough, so that quantum corrections become strong on the stretched horizon. In this model, the coupling strength\footnote{In this model the gravitational coupling is given by the value of $\left( e^{-2\phi}-\frac{\kappa}{4} \right)^{-\frac{1}{2}}$.} on the stretched horizon at the scrambling time is approximately given by $\left(M-\frac{\kappa}{4}t_S\right)^{-\frac{1}{2}}$. For instance, a ratio $M/\kappa=1$ yields a coupling strength of $\approx 1.2$ which indicates we should not trust the result at all for this choice of parameters. This is reflected in Figure~\ref{fig:cv_numeric_plots_a} which barely exhibits a linear growth rate. On the other hand a ratio $M/\kappa=100$ yields a coupling strength of $\approx0.1$ which demonstrates that we should be able to trust the solution for quite some time after the scrambling time. This is confirmed in Figure~\ref{fig:cv_numeric_plots_large_m} and \ref{fig:cv_numeric_plots_large_m_2} which shows a linearly falling growth rate over a long period of time.

\section{Conclusion and outlook}

In this paper we have computed holographic complexity as the volume of Einstein-Rosen bridges inside black holes
in two-dimensional dilaton gravity models where we have explicitly known semi-classical black hole solutions. This allows
us to follow the time evolution of the complexity of a black hole that is formed in gravitational collapse and subsequently
evaporates by emitting Hawking radiation.
Our main results can be summarized in three statements.

First, in order to obtain sensible results, we have to calculate extremal volumes in the four-dimensional parent theory
rather than lengths of geodesics in the two-dimensional reduction. The appropriately defined volume functional can be 
explicitly evaluated in the classical CGHS model and it exhibits the expected linear growth with Schwarzschild time at
late times. At the same time it is easy to check that the length of the spacelike geodesics, that will otherwise arise, does not lend itself to a direct 
interpretation in terms of complexity.

Second, when considering dynamical  black holes formed by gravitational collapse, we find it natural to cut off the 
volume integration at the stretched horizon of the black hole rather than extending the integration range to a distant
anchor curve. This distinction is unimportant if all we are interested in is the late time rate of growth of the complexity 
for a classical black hole but it does affect the onset of complexity growth. If the volume prescription extends to a distant 
anchor curve then the complexity already starts growing as soon the infalling shockwave passes the anchor point 
and the complexity growth turns on abruptly. If, on the other hand, we use the stretched horizon to delimit the integration
range, then complexity growth turns on smoothly at a time that coincides with the onset of Hawking emission at the
semi-classical level.

Third, using numerical methods, we obtain the complexity of an evaporating black hole as a function of time using the 
volume prescription inside the stretched horizon. We find that after the black hole is created, the complexity growth 
needs a time period of order the scrambling time to settle to a rate of growth proportional to the area of the stretched 
horizon. The growth rate then reliably tracks the area of the horizon as it shrinks due to black hole evaporation. 
Towards the end of the black hole lifetime, higher order quantum corrections are expected to become important
and semi-classical calculations can no longer be trusted. 

Holographic complexity can also be calculated in these models using the Wheeler-DeWitt action formalism.
We will present our results on that in a forthcoming companion paper~\cite{sst_inprep}, where we find that the 
numerical results obtained in the present paper for the volume complexity are confirmed by action calculations
that can be carried out analytically even at the semi-classical level.

Our results fit very well with Susskind's argument \cite{Susskind:2018pmk}, that the rate of complexity growth for
an evaporating black hole should at any given time be proportional to the product of the black hole entropy $S$ 
and the Hawking temperature $T$ at that time. In the models we are considering, the Hawking temperature 
remains constant and the entropy is proportional to black hole mass. It follows that the black hole loses area at
a constant rate determined by $N$ the number of matter channels available for Hawking emission. This translates 
into a growth rate of complexity that is initially proportional to the initial mass of the black hole and then drops 
linearly with time until the black hole has completely evaporated. In other words, $\dot{\mathcal{C}}\propto ST$ 
should decrease linearly in time. This precisely the behaviour we see in our numerical calculations, following an
initial onset period of order the scrambling time, which is logarithmic in $M$ in these models. 

Our results also support the notion that the holographic complexity corresponds to the quantum complexity
of the combined system of black hole and emitted Hawking radiation~\cite{Susskind:2018pmk}. The scrambling
dynamics that generates the growth in complexity takes place at the stretched horizon and the growth rate
is reduced as the area of the horizon shrinks. The outgoing radiation is free streaming and no further complexity
is generated by the degrees of freedom that have been emitted from the black hole. At the end of the day, the
black hole has disappeared and all that is left is a long train of outgoing radiation in a state of high, but no longer
growing, complexity.

\subsection*{Acknowledgments}
It is a pleasure to thank Shira Chapman and Nick Poovuttikul for discussions.
This research was supported by the Icelandic Research Fund under grants 185371-051 and 195970-051, and by the University of Iceland Research Fund.


\bibliographystyle{JHEP}
\bibliography{ref_comp_rst}

\providecommand{\href}[2]{#2}\begingroup\raggedright\begin{thebibliography}{10}

\bibitem{Susskind:1993if}
L.~Susskind, L.~Thorlacius and J.~Uglum, {\it {The Stretched horizon and black
  hole complementarity}},  {\em Phys. Rev.} {\bf D48} (1993) 3743--3761
  [\href{http://arXiv.org/abs/hep-th/9306069}{{\tt hep-th/9306069}}].

\bibitem{Susskind:2018pmk}
L.~Susskind, {\it {Three Lectures on Complexity and Black Holes}},  2018.
\newblock \href{http://arXiv.org/abs/1810.11563}{{\tt 1810.11563}}.

\bibitem{Susskind:2014rva}
L.~Susskind, {\it {Computational Complexity and Black Hole Horizons}},  {\em
  Fortsch. Phys.} {\bf 64} (2016) 44--48
  [\href{http://arXiv.org/abs/1403.5695}{{\tt 1403.5695}}]. [Fortsch.
  Phys.64,24(2016)].

\bibitem{Brown:2015bva}
A.~R. Brown, D.~A. Roberts, L.~Susskind, B.~Swingle and Y.~Zhao, {\it
  {Holographic Complexity Equals Bulk Action?}},  {\em Phys. Rev. Lett.} {\bf
  116} (2016), no.~19 191301 [\href{http://arXiv.org/abs/1509.07876}{{\tt
  1509.07876}}].

\bibitem{Brown:2015lvg}
A.~R. Brown, D.~A. Roberts, L.~Susskind, B.~Swingle and Y.~Zhao, {\it
  {Complexity, action, and black holes}},  {\em Phys. Rev.} {\bf D93} (2016),
  no.~8 086006 [\href{http://arXiv.org/abs/1512.04993}{{\tt 1512.04993}}].

\bibitem{sst_inprep}
L.~Schneiderbauer, W.~Sybesma and L.~Thorlacius, {\it {Action Complexity for
  Semi-Classical Black Holes}},  \href{http://arXiv.org/abs/2001.06453}{{\tt
  2001.06453}}.

\bibitem{Callan:1992rs}
C.~G. Callan, Jr., S.~B. Giddings, J.~A. Harvey and A.~Strominger, {\it
  {Evanescent black holes}},  {\em Phys. Rev.} {\bf D45} (1992), no.~4 R1005
  [\href{http://arXiv.org/abs/hep-th/9111056}{{\tt hep-th/9111056}}].

\bibitem{Russo:1992ht}
J.~G. Russo, L.~Susskind and L.~Thorlacius, {\it {Black hole evaporation in
  (1+1)-dimensions}},  {\em Phys. Lett.} {\bf B292} (1992) 13--18
  [\href{http://arXiv.org/abs/hep-th/9201074}{{\tt hep-th/9201074}}].

\bibitem{Giddings:1992kn}
S.~B. Giddings and A.~Strominger, {\it {Dynamics of extremal black holes}},
  {\em Phys. Rev.} {\bf D46} (1992) 627--637
  [\href{http://arXiv.org/abs/hep-th/9202004}{{\tt hep-th/9202004}}].

\bibitem{Banks:1992ba}
T.~Banks, A.~Dabholkar, M.~R. Douglas and M.~O'Loughlin, {\it {Are horned
  particles the climax of Hawking evaporation?}},  {\em Phys. Rev.} {\bf D45}
  (1992) 3607--3616 [\href{http://arXiv.org/abs/hep-th/9201061}{{\tt
  hep-th/9201061}}].

\bibitem{Harvey:1992xk}
J.~A. Harvey and A.~Strominger, {\it {Quantum aspects of black holes}},  in
  {\em {Proceedings, Theoretical Advanced Study Institute (TASI 92): From Black
  Holes and Strings to Particles: Boulder, USA, June 1-26, 1992}},
  pp.~537--588, 1993.
\newblock \href{http://arXiv.org/abs/hep-th/9209055}{{\tt hep-th/9209055}}.
\newblock [,122(1992)].

\bibitem{Thorlacius:1994ip}
L.~Thorlacius, {\it {Black hole evolution}},  {\em Nucl. Phys. Proc. Suppl.}
  {\bf 41} (1995) 245--275 [\href{http://arXiv.org/abs/hep-th/9411020}{{\tt
  hep-th/9411020}}].

\bibitem{Strominger:1994tn}
A.~Strominger, {\it {Les Houches lectures on black holes}},  in {\em {NATO
  Advanced Study Institute: Les Houches Summer School, Session 62: Fluctuating
  Geometries in Statistical Mechanics and Field Theory Les Houches, France,
  August 2-September 9, 1994}}, 1994.
\newblock \href{http://arXiv.org/abs/hep-th/9501071}{{\tt hep-th/9501071}}.

\bibitem{Giddings:1994pj}
S.~B. Giddings, {\it {Quantum mechanics of black holes}},  in {\em
  {Proceedings, Summer School in High-energy physics and cosmology: Trieste,
  Italy, June 13-July 29, 1994}}, pp.~0530--574, 1994.
\newblock \href{http://arXiv.org/abs/hep-th/9412138}{{\tt hep-th/9412138}}.

\bibitem{Brown:2018bms}
A.~R. Brown, H.~Gharibyan, H.~W. Lin, L.~Susskind, L.~Thorlacius and Y.~Zhao,
  {\it {Complexity of Jackiw-Teitelboim gravity}},  {\em Phys. Rev.} {\bf D99}
  (2019), no.~4 046016 [\href{http://arXiv.org/abs/1810.08741}{{\tt
  1810.08741}}].

\bibitem{Chapman:2018dem}
S.~Chapman, H.~Marrochio and R.~C. Myers, {\it {Holographic complexity in
  Vaidya spacetimes. Part I}},  {\em JHEP} {\bf 06} (2018) 046
  [\href{http://arXiv.org/abs/1804.07410}{{\tt 1804.07410}}].

\bibitem{Bilal:1992kv}
A.~Bilal and C.~G. Callan, Jr., {\it {Liouville models of black hole
  evaporation}},  {\em Nucl. Phys.} {\bf B394} (1993) 73--100
  [\href{http://arXiv.org/abs/hep-th/9205089}{{\tt hep-th/9205089}}].

\bibitem{deAlwis:1992emy}
S.~P. de~Alwis, {\it {Quantization of a theory of 2-d dilaton gravity}},  {\em
  Phys. Lett.} {\bf B289} (1992) 278--282
  [\href{http://arXiv.org/abs/hep-th/9205069}{{\tt hep-th/9205069}}].

\end{thebibliography}\endgroup
\end{document}